\documentclass[12pt]{article}
\usepackage{amsfonts}

\usepackage{graphicx}
\usepackage{amsmath}
\usepackage{float}

\textwidth=6.5in \hoffset=-.75in \voffset=-1in \numberwithin{equation}{section}
\numberwithin{figure}{section}

\input{tcilatex}

\begin{document}

\begin{titlepage}
\bigskip \begin{flushright}
\end{flushright}
\vspace{1cm}
\begin{center}
{\Large \bf {Resolved Conifolds in Supergravity Solutions}}\\
\end{center}
\vspace{2cm}
\begin{center}
 A. M.
Ghezelbash{ \footnote{ EMail: amasoud@sciborg.uwaterloo.ca}}
\\
Department of Physics and Astronomy, University of Waterloo, \\
Waterloo, Ontario N2L 3G1, Canada\\
\vspace{1cm}
\end{center}

\begin{abstract}

We construct generalized 11D supergravity solutions of fully localized intersecting 
D2/D4 brane
systems. These solutions are obtained by embedding six-dimensional 
resolved Eguchi-Hanson conifolds lifted to M-theory. We reduce these
solutions to ten dimensions, obtaining new D-brane systems in type IIA supergravity.
We discuss the limits in which the dynamics of the D2 brane decouples from the bulk for these solutions. 

\end{abstract}
\bigskip
\end{titlepage}\onecolumn

\bigskip

\section{Introduction}

\bigskip The correspondence between field theories in $(d+1)-$dimensional
Anti-de Sitter space-time and $d-$dimensional conformal field theories has
been studied in various aspects in the last years \cite{Mal}-\cite{Chal}. \
The correspondence has been conjectured for large $N$ limit of
superconformal gauge theory on the boundary of $AdS$ and the supergravity on
the bulk of $AdS_{\text{ }}$\cite{Mal}. The conjecture states that partition
function of the field theory on the bulk of $AdS$ should be identified with
the generating functional of the boundary superconformal field theory.
Different aspects of correspondence in string theory on $AdS_{5}\times S^{5}$
have been tested and carried out in \cite{Chal}-\cite{ADS}. \ The
correspondence also has been considered and studied on the bulk/boundary of $%
AdS_{5}\times \mathcal{E}$ where $\mathcal{E}$ is Sasaki-Einstein manifold.
This means that the metric cone $\mathcal{E}$ is Ricci-flat K\"{a}hler or
Calabi-Yau where the superconformal field theory may be arising from
D3-branes sitting at the tip of Calabi-Yau cone. \ By deforming the tip of
Calabi-Yau cones, one can obtain resolved Calabi-Yau manifolds with smooth
geometries \cite{Candelas}.

In a related line of research, recently new supergravity solutions for fully
localized intersecting type IIA D2$\perp $D6, D2$\perp $D4, NS5$\perp $D6
and NS5$\perp $D5 brane systems have been obtained \cite{hashi,CGMM2,ATM2}.
By lifting a D6 (D5 or D4)-brane to four-dimensional Bianchi type IX
geometry embedded in M-theory \cite{Me}, these solutions were constructed by
placing M2- and M5-branes in the four-dimensional Taub-NUT/Bolt,
Eguchi-Hanson and Atiyah-Hitchin background geometries (all as special cases
of the four-dimensional Bianchi type IX geometry) as well as Bianchi type IX
geometry. The special feature of these constructions\ is that the solution
is not restricted to be in the near core region of the D6 (D5 or D4)-brane.\
\ 

Inspired with these works, in this paper we embed the six-dimensional
Eguchi-Hanson resolved conifold which has been constructed first in \cite{Z}
into M-theory.\ Although some higher-dimensional Eguchi-Hanson metrics have
been constructed in \cite{highereh}, but these solutions are asymptotically
(A)dS and couldn't be embedded in M-theory. The solutions in the paper do
not preserve any supersymmetry due to dimension of embedded Eguchi-Hanson
space, but nevertheless exhibit interesting properties that are
qualitatively similar to supersymmetric solutions \cite{hashi,CGMM2,ATM2}.
Specifically the brane metric function behaves the same way near the brane
core and at infinity. Moreover it is an integrated product of a decaying
function and a damped oscillating function far from the brane. Near the
brane core, the convolution of the two functions diverges as for the
supersymmetric cases. For all of the different solutions we compactify the
solutions on a circle, obtaining the different fields of type IIA string
theory. Explicit calculation shows that in all cases the metric is
asymptotically (locally) flat, though for some of the compactified solutions
the type IIA dilaton field diverges at infinity. \ The outline of this paper
is as follows. In section \ref{sec:review}, we discuss briefly the
six-dimensional Eguchi-Hanson conifolds. In section \ref{sec:M2}, we present
the different M2-brane solutions on the background of resolved Eguchi-Hanson
conifolds and find type IIA D2$\perp $D4(2) intersecting brane solutions
upon dimensional reduction. \ In section \ref{sec:declim}, we consider the
decoupling limit of these solutions.

\section{Eguchi-Hanson Resolved Conifolds}

\label{sec:review}

\bigskip The four-dimensional Eguchi-Hanson metric is given by 
\begin{equation}
ds^{2}=g(r)dr^{2}+\frac{r^{2}}{4g(r)}[d\psi +\cos (\theta )d\phi ]^{2}+\frac{%
r^{2}}{4}(d\theta ^{2}+\sin ^{2}(\theta )d\phi ^{2}),  \label{4DEH}
\end{equation}%
where 
\begin{equation}
g(r)=(1-\frac{a^{4}}{r^{4}})^{-1}.  \label{g4DEH}
\end{equation}%
The space is asymptotically locally Euclidean with a self-dual curvature.
The metric has a single removable bolt singularity if $\psi $ is restricted
to the interval $(0,2\pi )$ and the topology of the manifold is $S^{3}/Z_{2}$
asymptotically, hence the manifold is asymptotically locally Euclidean.
Moreover the manifold has the topology of $R^{2}\times S^{2}$ near bolt
singularity $r=a.$

The six-dimensional Eguchi-Hanson class of instantons on the conifold was
first explicitly obtained in \cite{Z}. The solution was found explicitly by
constructing the Ricci-flat K\"{a}hler metric on resolved conifold. \ The
metric is%
\begin{equation}
ds_{EH_{6}}^{2}=f(r)dr^{2}+\frac{r^{2}}{9f(r)}[d\psi +\cos (\theta
_{1})d\phi _{1}+\cos (\theta _{2})d\phi _{2}]^{2}+\frac{r^{2}}{6}(d\theta
_{1}^{2}+\sin ^{2}(\theta _{1})d\phi _{1}^{2}+d\theta _{2}^{2}+\sin
^{2}(\theta _{2})d\phi _{2}^{2}),  \label{6DEH}
\end{equation}%
where the metric function $f(r)$ is given by $f(r)=(1-\frac{a^{6}}{r^{6}}%
)^{-1}.$ We note that the metric (\ref{6DEH}) was obtained in \cite{Rob} as $%
t-$constant hypersurface of the seven dimensional Eguchi-Hanson-AdS (dS)
solitons in the limit of zero cosmological constant.

We note that for $a=0$, the metric (\ref{6DEH}) reduces to metric of
Calabi-Yau cone $ds^{2}=dr^{2}+r^{2}d\Sigma _{1,1}^{2}$ where $d\Sigma
_{1,1}^{2}$ is the metric on manifold $\mathcal{N}_{1,1\text{ \ }}$which is
fibre bundle over $S^{2}\times S^{3}.$ In general the metrics $d\Sigma
_{p,q}^{2}$ given by%
\begin{equation}
d\Sigma _{p,q}^{2}=\alpha \lbrack d\psi +p\cos (\theta _{1})d\phi _{1}+q\cos
(\theta _{2})d\phi _{2}]^{2}+\beta _{1}(d\theta _{1}^{2}+\sin ^{2}(\theta
_{1})d\phi _{1}^{2})+\beta _{2}(d\theta _{2}^{2}+\sin ^{2}(\theta _{2})d\phi
_{2}^{2}),  \label{c}
\end{equation}%
in which $p$ and $q$ are relatively prime integers, are the metrics on
manifolds $\mathcal{N}_{p,q}^{\text{ }}$ which are $U(1)$ fibre bundles over 
$S^{2}\times S^{2}$. The coordinates $\theta _{i},\phi _{i}$ are spherical
polar coordinates on each $S^{2}$ and $\psi $ is the coordinate on the $U(1)$
fibre. For two special choices of $p=q=1$ and $p=1,q=0$ these fibre bundles
are over $S^{2}\times S^{3}.$ In the first case $\alpha =\frac{1}{9},\beta
_{1}=\frac{1}{6},\beta _{2}=\frac{1}{6}$ and the metric of Calabi-Yau cone $%
ds^{2}=dr^{2}+r^{2}d\Sigma _{1,1}^{2}$ is an special case of six-dimensional
Eguchi-Hanson instantons (\ref{6DEH}) and in the second case $\alpha =\beta
_{1}=\frac{1}{8},\beta _{2}=\frac{1}{4}$ with no special relation to the
metric (\ref{6DEH}). Although two manifolds $\mathcal{N}_{1,1\text{ \ }}$and 
$\mathcal{N}_{1,0\text{ \ }}$are diffeomorphic but two metrics $d\Sigma
_{1,1}^{2}$ and $d\Sigma _{1,0}^{2}$ represent different geometries on $%
S^{2}\times S^{3}.$

\section{Embedding Six-dimensional Eguchi-Hanson Resolved Conifolds in M2
brane}

\bigskip \label{sec:M2}

We embed the six-dimensional Eguchi-Hanson metric explicitly into the
eleven-dimensional supergravity metric given by 
\begin{equation}
ds_{11}^{2}=H^{-2/3}(y,r)(-dt^{2}+dx_{1}^{2}+dx_{2}^{2})+H^{1/3}(y,r)(dy^{2}+y^{2}d\alpha ^{2}+ds_{EH6}^{2}),
\label{ds11TN6}
\end{equation}%
with the following non-vanishing components of four-form field $F,$ 
\begin{equation}
F_{tx_{1}x_{2}y}=-\frac{1}{2H^{2}}\frac{\partial H}{\partial y},
\label{Ft12yTN6}
\end{equation}

\begin{equation}
F_{tx_{1}x_{2}r}=-\frac{1}{2H^{2}}\frac{\partial H}{\partial r}.
\label{Ft12rTN6}
\end{equation}%
We note that to construct a solution to the equations of 11-dimensional
supergravity and successfully reduced them to $D=10$ dimensional type IIA
string theory, we must assume a bosonic ground state, i.e. the vacuum
expectation value of any fermionic field should be zero. This will allow us
to focus on the equations for $g_{MN}$ and $A_{MNP}$ $(M,N,P=0,...,10),$
which are now given by 
\begin{eqnarray}
R_{MN}-\frac{1}{2}g_{MN}R &=&\frac{1}{3}\left[ F_{MPQR}F_{N}^{~PQR}-\frac{1}{%
8}g_{MN}F_{PQRS}F^{PQRS}\right] ,  \label{GminG2} \\
\nabla _{M}F^{MNPQ} &=&-\frac{1}{576}\varepsilon ^{M_{1}\ldots
M_{8}NPQ}F_{M_{1}\ldots M_{4}}F_{M_{5}\ldots M_{8}},  \label{dFgen}
\end{eqnarray}%
where because $\left\langle \Psi _{M}\right\rangle =0$, $F_{MNPQ}$ is the
unmodified four-form field strength 
\begin{eqnarray}
F_{MNPQ} &=&4\partial _{\lbrack M}A_{NPQ]}  \notag \\
&=&\frac{1}{2}\left[ A_{MNP,Q}-A_{NPQ,M}+A_{PQM,N}-A_{QMN,P}\right] .
\label{Fgen}
\end{eqnarray}%
{\large \ }From (\ref{ds11TN6}), one can use 
\begin{equation}
g_{AB}=\left[ 
\begin{array}{cc}
e^{-2\Phi /3}\left( g_{\alpha \beta }+e^{2\Phi }C_{\alpha }C_{\beta }\right)
& \nu e^{4\Phi /3}C_{\alpha } \\ 
\nu e^{4\Phi /3}C_{\beta } & \nu ^{2}e^{4\Phi /3}%
\end{array}%
\right] ,  \label{genkkmetric}
\end{equation}%
where $\nu $ is the winding number, giving the number of times the membrane
wraps around the compactified dimension \cite{Townsend}. For simplicity we
will take $\nu =1$\ in what follows. From (\ref{genkkmetric}) and the
reduction of $A_{MNP}$ to its ten dimensional form, the Ramond-Ramond (RR) ($%
C_{\alpha }$, $A_{\alpha \beta \gamma }$) and Neveu-Schwarz Neveu-Schwarz
(NSNS) ($\Phi $, $B_{\alpha \beta }$ and $g_{\alpha \beta }$) fields can be
easily read off. Once the ten dimensional equations are found, their
analysis and comparison to existing forms can be carried out. In obtaining
the relation (\ref{genkkmetric}) with $\nu =1$, we use the well known
Kaluza-Klein reduction of the 11D supergravity metric and field strength to
10D metric and field strength \cite{SMITH} 
\begin{eqnarray}
ds_{(1,10)}^{2} &=&e^{-2\Phi /3}ds_{(1,9)}^{2}+e^{4\Phi
/3}(dx_{10}+C_{\alpha }dx^{\alpha })^{2},  \label{KKred} \\
F_{(4)} &=&\mathcal{F}_{(4)}+H_{(3)}\wedge dx_{10},  \label{KKred2}
\end{eqnarray}%
where $\mathcal{F}_{(4)}$ and $H_{(3)}$\ are the RR four-form and NSNS
three-form field strengths corresponding to $A_{\alpha \beta \gamma }$ and $%
B_{\alpha \beta }$ and $x_{10}$\ is the coordinate of compactified manifold.
We take it to be a circle with radius $R_{\infty }$, parameterized as $%
x_{10}=R_{\infty }\psi $\ where $\psi $\ has period $2\pi .$\ Although we
have assumed $\nu =1,$ \ the $\nu \neq 1$ case can be dealt with by
compactifying $\nu $\ times over this circle and replacing $x_{10}$\ by $\nu
x_{10}$\ in the relations (\ref{KKred}) and (\ref{KKred2}). This simply adds
to the dilaton field a constant term of the form $\frac{3}{2}\ln \nu ,$\ and
multiplies the RR field $C_{\alpha }$ by a multiplicative constant of $\frac{%
1}{\nu }$ when we reduce the theory to 10 dimensions.

Requiring that (\ref{ds11TN6}), (\ref{Ft12yTN6}) and (\ref{Ft12rTN6})
satisfy the field equations (\ref{GminG2}) and (\ref{dFgen}) yields the
differential equation 
\begin{equation}
r^{7}\frac{\partial ^{2}H(y,r)}{\partial y^{2}}+r^{7}\frac{\partial H(y,r)}{%
y\partial y}+r(r^{6}-a^{6})\frac{\partial ^{2}H(y,r)}{\partial r^{2}}%
+(5r^{6}+a^{6})\frac{\partial H(y,r)}{\partial r}=0,  \label{diffeqtn4}
\end{equation}%
for $H(y,r)$. By substituting 
\begin{equation}
H(y,r)=1+Q_{M2}R(r)Y(y),  \label{Hyrsoln}
\end{equation}%
where $Q_{M2}$ is the M2 brane charge, we obtain 
\begin{equation}
\frac{\partial ^{2}Y}{\partial y^{2}}+\frac{1}{y}\frac{\partial Y}{\partial y%
}-c^{2}Y=0,  \label{BesselJeqn}
\end{equation}%
whose solution is $K_{0}(cy)$ which is decaying to zero at large $y.$ The
differential equation for $R(r)$ turns out to be%
\begin{equation}
r(r^{6}-a^{6})\frac{d^{2}R}{dr^{2}}+(5r^{6}+a^{6})\frac{dR}{dr}%
+c^{2}r^{7}R(r)=0.  \label{hashiHyr}
\end{equation}%
We solve the radial differential equation (\ref{hashiHyr}) in two different
cases:

\textbf{I)} $a=0$; in this case the transverse space to the M2 brane is
product of \ $\mathbb{R}^{2}$\ with Calabi-Yau cone $ds^{2}=dr^{2}+r^{2}d%
\Sigma _{1,1}^{2}$. The solution to equation (\ref{hashiHyr}) is given by%
\begin{equation}
R(r)\sim \frac{1}{r^{2}}J_{2}(cr),  \label{Rsol}
\end{equation}%
where \ $J_{2}$ is the first order Bessel function of the second kind. We
note that solution (\ref{Rsol}) is finite for $r$ near to and decays away
from the tip of Calabi-Yau cone. Hence, the most general form for the metric
function is given by%
\begin{equation}
H(y,r)=1+Q_{M2}\int_{0}^{\infty }dcf(c)\frac{1}{r^{2}}J_{2}(cr)K_{0}(cy).
\label{HEH}
\end{equation}%
To fix up the measure function $f(c)$, we should consider the near horizon
limit of our solution. In this limit where $r$ approaches the tip of the
cone, the transverse space to brane reduces to $\mathbb{R}^{2}\times
S^{3}\times S^{3}$ and hence the metric function should coincide with $1+%
\frac{Q_{M2}}{y^{6}}$. By comparing the metric function (\ref{HEH}) in near
horizon limit with the known integral \ 
\begin{equation}
\int_{0}^{\infty }dcc^{5}K_{0}(cy)=(\frac{2}{y})^{6},  \label{Kint}
\end{equation}%
we obtain $f(c)=\frac{c^{3}}{8}$ and so the metric function becomes%
\begin{equation}
H(y,r)=1+\frac{Q_{M2}}{8}\int_{0}^{\infty }c^{3}dc\frac{1}{r^{2}}%
J_{2}(cr)K_{0}(cy)=1+\frac{Q_{M2}}{(y^{2}+r^{2})^{3}}.  \label{Hazero}
\end{equation}%
So we notice that although the transverse space to the M2 brane has conical
singularity at $r=0$, but M2 brane metric function (\ref{Hazero}) has proper
behaviour over there. This feature is quite interesting since all the other
previously know solutions \cite{hashi,CGMM2,ATM2} are absolutely free of
conical singularities in the transverse spaces.

\bigskip

\begin{figure}[tbp]
\centering                 
\begin{minipage}[c]{.3\textwidth}
        \centering
        \includegraphics[width=\textwidth]{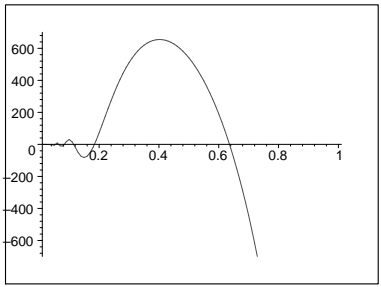}
    \end{minipage}
\caption{Numerical solution $R/10^{5}$ to radial equation (\ref{hashiHyr})
as a function of $\frac{a}{r}.$ For $r=a$, the radial function $R$ diverges
and for $r\rightarrow \infty $, it vanishes as $\frac{J_{2}(cr)}{r^{2}}$. }
\label{fig1}
\end{figure}

\textbf{II) } $a\neq 0$; In this case, we can find the solutions to equation
(\ref{hashiHyr}) numerically since it is not solvable analytically (unless $%
c=0$, which reduces generality). For large $r$, the solution to equation (%
\ref{hashiHyr}) that vanishes at infinity is $\frac{J_{2}(cr)}{r^{2}}$. A
typical numerical solution to the radial differential equation versus $a/r$
is displayed in figure \ref{fig1}. So the metric function becomes%
\begin{equation}
H_{EH}(y,r)=1+Q_{M2}\int_{0}^{\infty }dcp(c)R(r)K_{0}(cy).  \label{Hanonzero}
\end{equation}%
By dimensional analysis, $p(c)=p_{0}c^{5}$, where $p_{0}$\ is a constant
that can be absorbed into the definition of $Q_{M2}$. We notice the radial
part of the metric function approaches to the finite value of $\frac{c^{2}}{8%
}$ on the tip of the cone. On the other hand by resolving the singularity of
the cone in the transverse space to $M2$ brane, the radial part of the
metric function diverges near the point $r=a$ (we note that in metric (\ref%
{6DEH}), the coordinate $r$ is greater than or equal to $a$).

\bigskip By changing $c\rightarrow ic$ in the differential equations (\ref%
{BesselJeqn}) and (\ref{hashiHyr}), we get another solution in the form of 
\begin{equation}
\widetilde{H}_{EH}(y,r)=1+Q_{M2}\int_{0}^{\infty }dcc^{5}\widetilde{R}%
_{c}(r)Y_{0}(cy)  \label{HEH4sc}
\end{equation}%
where $Y_{0}(cy)$\ is the Bessel function of second kind and $\widetilde{R}%
_{c}(r)$ is the solution to%
\begin{equation}
r(r^{6}-a^{6})\frac{d^{2}\widetilde{R}_{c}(r)}{dr^{2}}+(5r^{6}+a^{6})\frac{d%
\widetilde{R}_{c}(r)}{dr}-c^{2}r^{7}\widetilde{R}_{c}(r)=0.  \label{Rtildeeq}
\end{equation}

\begin{figure}[tbp]
\centering                 
\begin{minipage}[c]{.3\textwidth}
        \centering
        \includegraphics[width=\textwidth]{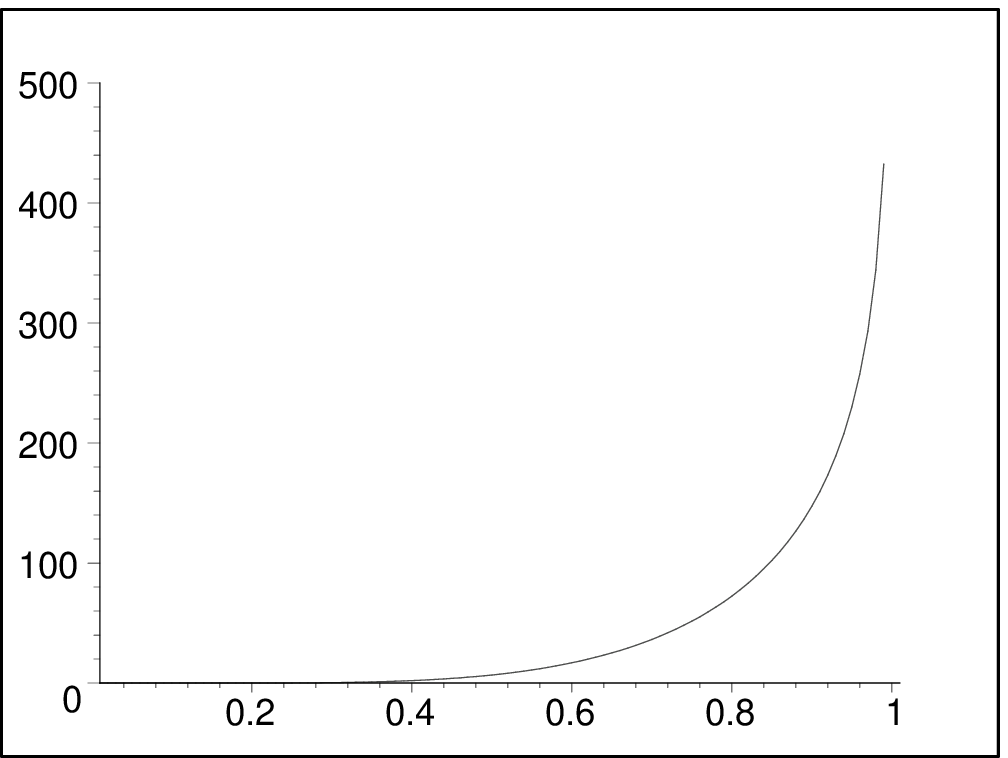}
    \end{minipage}
\caption{Numerical solution $\widetilde{R}/10^{5}$ to radial equation (\ref%
{Rtildeeq}) as a function of $\frac{a}{r}.$ For $r=a$, the radial function $%
R $ diverges and for $r\rightarrow \infty $, it vanishes as $\frac{e^{-cr}}{%
r^{5/2}}$. }
\label{fig2}
\end{figure}

At large $r,$ the radial solution monotonically vanishes as $\frac{e^{-cr}}{%
r^{5/2}}.$ A typical numerical solution to the equation (\ref{Rtildeeq})
versus $a/r$ is displayed in figure \ref{fig2}. 

Reduction of these solutions to a ten dimensional type IIA string theory
solution proceeds in a manner similar to the reduction of $S^{4}$ regarded
as containing a NUT and anti-NUT charge \cite{ChamGib}. \ The vector $%
\partial /\partial \psi $ generates a Hopf fibration of a 3-sphere in the
metric (\ref{6DEH}). Using (\ref{genkkmetric}) we obtain the NSNS fields%
\begin{equation}
\begin{array}{rcl}
\Phi & = & \frac{3}{4}\ln \left\{ \frac{H^{1/3}w^{2}}{9f}\right\} \\ 
B_{\mu \nu } & = & 0%
\end{array}
\label{NSNSEH}
\end{equation}%
where we define the dimensionless coordinate $w$ by $w=\frac{r}{a}$. The
Ramond-Ramond (RR) fields and the ten dimensional metric will be given by 
\begin{equation}
\begin{array}{rcl}
C_{\phi _{i}} & = & a\cos (\theta _{i}) \\ 
A_{tx_{1}x_{2}} & = & \frac{1}{H}%
\end{array}%
\end{equation}%
\begin{eqnarray}
ds_{10}^{2} &=&\frac{w}{3}\{H^{-1/2}f^{-1/2}\left(
-dt^{2}+dx_{1}^{2}+dx_{2}^{2}\right) +  \notag \\
&+&H^{1/2}f^{-1/2}\left( dy^{2}+y^{2}d\alpha ^{2}\right)
+H^{1/2}f^{1/2}a^{2}\times  \notag \\
&\times &[dw^{2}+\frac{w^{2}}{6f}(d\theta _{1}^{2}+\sin ^{2}(\theta
_{1})d\phi _{1}^{2}+d\theta _{2}^{2}+\sin ^{2}(\theta _{2})d\phi
_{2}^{2})]\}.  \label{dsEH10}
\end{eqnarray}%
The metric (\ref{dsEH10}) describes a D2$\perp $D4 system where the D2-brane
is localized along the world-volume of the D4-brane. We note $H$ refers to
either $H_{EH}$ or $\widetilde{H}_{EH}$ given by (\ref{Hanonzero}) and (\ref%
{HEH4sc}) and $f=f(w)=\left( 1-w^{-6}\right) ^{-1}$ in the above equations.
We have explicitly checked that the above 10-dimensional metric, with the
given dilaton and one form, is a solution to the 10-dimensional supergravity
equations of motion.

The 10D metric is locally asymptotically flat (though the dilaton field
diverges); for large $w$ it\ reduces to 
\begin{equation}
ds_{10}^{2}=\frac{w}{3}\{-dt^{2}+dx_{1}^{2}+dx_{2}^{2}+dy^{2}+y^{2}d\alpha
^{2}+a^{2}[dw^{2}+\frac{w^{2}}{6}(d\Omega _{2}^{2}+d\Omega _{2}^{\prime
2})]\}  \label{ds10Ehlarger}
\end{equation}%
which is a 10D locally flat metric with solid deficit angles. The Kretchmann
invariant of this spacetime vanishes at infinity and is given by 
\begin{equation}
R_{\mu \nu \rho \sigma }R^{\mu \nu \rho \sigma }=\frac{2214}{w^{6}}
\label{KEHlarger}
\end{equation}%
and all the components of the Riemann tensor in the orthonormal basis have
similar $\frac{1}{w^{3}}$ behavior, vanishing at infinity.

\section{Decoupling limits}

\bigskip \label{sec:declim}

At low energies, the dynamics of the D2 brane decouple from the bulk, with
the region close to the D4 brane corresponding to a range of energy scales
governed by the IR fixed point \cite{DecouplingLim}. For D2 branes localized
on D4 branes, this corresponds in the field theory to a vanishing mass for
the fundamental hyper-multiplets. Near the D2 brane horizon ($H\gg 1$), the
field theory limit is given by 
\begin{equation}
g_{YM2}^{2}=g_{s}\ell _{s}^{-1}=\text{fixed.}  \label{gymFTlimit}
\end{equation}%
In this limit the gauge coupling on the four brane $g_{YM4}^{2}=(2\pi
)^{2}g_{s}\ell _{s}=(2\pi \ell _{s})^{2}g_{YM2}^{2}$ goes to zero, so the
dynamics there decouple. In each of our cases above, the radial coordinates
are also scaled such that 
\begin{equation}
Y=\frac{y}{\ell _{s}^{2}}~~,~~~~U=\frac{r}{\ell _{s}^{2}}
\label{YUdecoupling1}
\end{equation}%
are fixed. \ As a result, we note that this will change the harmonic
function of the D6 brane and we get%
\begin{equation}
f(r)\rightarrow \left( 1-\frac{A^{6}}{U^{6}}\right) =f(U)  \label{EHfU}
\end{equation}%
where $a$ has been rescaled to $a=A\ell _{s}^{2}$. The radial functions $R$
and $\widetilde{R}$ in equations (\ref{Hanonzero}) and (\ref{HEH4sc}) are
the solutions of\bigskip

\begin{equation}
U(U^{6}-A^{6})\frac{d^{2}R(U)}{dU^{2}}+(5U^{6}+A^{6})\frac{dR(U)}{dU}%
+C^{2}U^{7}R(U)=0  \label{cc}
\end{equation}%
\begin{equation}
U(U^{6}-A^{6})\frac{d^{2}\widetilde{R}(U)}{dU^{2}}+(5U^{6}+A^{6})\frac{d%
\widetilde{R}(U)}{dU}-C^{2}U^{7}\widetilde{R}(U)=0  \label{ff}
\end{equation}%
respectively, where we rescaled the integration variable $c$ by $C/\ell
_{s}^{2}$. \ Moreover, we use $\ell _{p}=g_{s}^{1/3}\ell _{s}$ to rewrite 
\begin{equation}
Q_{M2}=32\pi ^{2}N_{2}\ell _{p}^{6}=32\pi ^{2}N_{2}g_{YM2}^{2}\ell _{s}^{8}
\label{Qm2value}
\end{equation}%
and so the M2 metric function (\ref{Hanonzero}) changes to%
\begin{equation}
H_{EH}(y,r)\rightarrow Q_{M2}\int \frac{dC}{\ell _{s}^{2}}p_{0}\frac{C^{5}}{%
\ell _{s}^{10}}R(U)K_{0}(CY)=\frac{1}{\ell _{s}^{4}}h_{EH}(Y,U)  \label{ffk}
\end{equation}

The other D2 harmonic function in the above solution (\ref{HEH4sc}) can be
shown to scale as $\widetilde{H}_{EH}(Y,U)=\ell _{s}^{-4}\widetilde{h}%
_{EH}(Y,U)$ in the same was as $H_{EH}(y,r).$ This scaling form causes the
D2-brane to warp the ALE region and the asymptotically flat region of the
D4-brane geometry.

\ Finally, the ten-dimensional metric (\ref{dsEH10}) scales as\bigskip 
\begin{eqnarray}
\frac{ds_{10}^{2}}{\ell _{s}^{2}} &=&\frac{U}{3A}\{h^{-1/2}(Y,U)f^{-1/2}(U)%
\left( -dt^{2}+dx_{1}^{2}+dx_{2}^{2}\right) +  \notag \\
&+&h^{1/2}(Y,U)f^{-1/2}(U)\left( dY^{2}+Y^{2}d\alpha ^{2}\right)
+h^{1/2}(Y,U)f^{1/2}(U)\times  \notag \\
&\times &[dU^{2}+\frac{U^{2}}{6f(U)}(d\theta _{1}^{2}+\sin ^{2}(\theta
_{1})d\phi _{1}^{2}+d\theta _{2}^{2}+\sin ^{2}(\theta _{2})d\phi
_{2}^{2})]\}.
\end{eqnarray}%
and there is only an overall normalization factor of $\ell _{s}^{2}$ in the
above metric which is the expected result for a solution that is a
supergravity dual of a quantum field theory.

\section{Conclusions}

By embedding six-dimensional Eguchi-Hanson resolved conifolds into M-theory,
we have found new classes of \ 2-brane solutions to $D=11$\ supergravity.
These exact solutions are new M2- brane metrics with metric functions (\ref%
{Hanonzero}), (\ref{HEH4sc})- these are the main results of this paper. The
conical singularity of Eguchi-Hanson geometry makes this space special to be
considered as an interesting transverse space to M2-brane. The brane
solutions show that even in the presence of conical singularities in the
transverse space (unlike the other know solutions that are free of conical
singularities in the transverse space), we can find a brane metric function.
The common feature of solutions is that the brane function is a convolution
of an decaying `radial' function with a damped oscillating one. The `radial'
function vanishes far from the branes and diverges near the brane core,
while without resolving the cone singularity, the radial function approaches
a finite value on the tip of the cone. Finally we considered the decoupling
limit of our solutions and found that all solutions behave properly in the
decoupling limit.

\bigskip

{\Large Acknowledgments}

I would like to thank Robert Mann for discussion.


\bigskip

\begin{thebibliography}{99}
\bibitem{Mal} J.M. Maldacena, \textit{Adv. Theor. Math. Phys.} \textbf{2}
(1998) 231.

\bibitem{Lu} H. Liu and A. A. Tseytlin, \textit{Nucl. Phys.} \textbf{B533}
(1998) 88.

\bibitem{BanksG} T. Banks and M. B. Green, \textit{JHEP}\textbf{\ 9805}
(1998) 002.

\bibitem{Chal} G. Chalmers, H. Nastase, K. Schalm and R. Siebelink, \textit{%
Nucl. Phys.} \textbf{B540} (1999) 247.

\bibitem{ADS} A.M. Ghezelbash, K. Kaviani, S. Parvizi and A.H. Fatollahi, 
\textit{Phys. Lett.} \textbf{B435} (1998) 291.

\bibitem{Candelas} P. Candelas and X.C. de la Ossa, \textit{Nucl. Phys.} 
\textbf{B342} (1990) 246.

\bibitem{hashi} S.A. Cherkis and A. Hashimoto, \textit{JHEP}\textbf{\ 0211 }%
(2002) 036.

\bibitem{CGMM2} R. Clarkson, A.M. Ghezelbash and R.B. Mann, \textit{JHEP} 
\textbf{0404} (2004) 063; \textbf{0408} (2004) 025.

\bibitem{ATM2} A.M. Ghezelbash and R.B. Mann, \textit{JHEP} \textbf{0410}
(2004) 012.

\bibitem{Me} A.M. Ghezelbash, \textit{Phys.Rev.} \textbf{D74} (2006) 126004.

\bibitem{Z} L.A. Pando Zayas, A.A. Tseytlin, \textit{Phys.Rev.} \textbf{D63}
(2001) 086006 .

\bibitem{highereh} R. Clarkson and R.B. Mann, \textit{Phys.Rev.Lett. }%
\textbf{96 }(2006)\textbf{\ }051104; R. B. Mann, C. Stelea, \textit{%
Phys.Lett. }\textbf{B634} (2006) 448.

\bibitem{Rob} R. Clarkson and R.B. Mann, \textit{Class.Quant.Grav.} \textbf{%
23} (2006) 1507.

\bibitem{Townsend} P.K. Townsend, hep-th/9612121.

\bibitem{SMITH} D.J. Smith, \textit{Class.Quant.Grav.} \textbf{20} (2003)
R233.

\bibitem{ChamGib} R.R. Caldwell, H. A. Chamblin, and G.W. Gibbons, \textit{%
Phys. Rev.} \textbf{D53} (1996) 7103.

\bibitem{DecouplingLim} N. Itzhaki, A.A. Tseytlin and S. Yankielowicz, 
\textit{Phys.Lett.} \textbf{B432} (1998) 298; J. M. Maldacena, \textit{\
Adv.Theor.Math.Phys.} \textbf{2} (1998) 231; \textit{Int.J.Theor.Phys.} 
\textbf{38} (1999) 1113; N. Itzhaki, J.M. Maldacena, J. Sonnenschein and S.
Yankielowicz, \textit{Phys.Rev.} \textbf{D58} (1998) 046004.
\end{thebibliography}
\end{document}